\documentclass[letterpaper,twocolumn,10pt]{article}
\usepackage[
  pass,%
]{geometry}
\newcount\Comments  \Comments=0   %

\usepackage{array,multirow}
\usepackage{graphicx} 
\usepackage[usenames, dvipsnames]{color} 
\usepackage{enumitem} 
\usepackage{amsmath}
\usepackage{amssymb}
\usepackage{cite}
\usepackage{booktabs}
\usepackage{amssymb}
\usepackage{xspace}
\usepackage{times}
\usepackage{balance}
\usepackage{pifont}
\usepackage{xcolor}
\usepackage[small,bf]{caption}

\usepackage[hyphens]{url}

\newcommand{\kibitz}[2]{\ifnum\Comments=1\textcolor{#1}{#2}\fi}
\newcommand{\sbk}[1]{\kibitz{Green}{[\textbf{Bano} -- \emph{#1}]}}
\newcommand{\nv}[1]{\kibitz{Red}{[\textbf{Narseo} -- \emph{#1}]}}

\newcommand{\edc}[1]{\kibitz{purple}{[\textbf{EdC} -- \emph{#1}]}}

\def\first{({\it i})\xspace}
\def\second{({\it ii})\xspace}
\def\third{({\it iii})\xspace}

\def\yes{\checkmark}
\def\no{\ding{56}}

\def\UrlFont{\small\ttfamily}

\providecommand{\adblocker}{adblocker\xspace}
\providecommand{\adblockers}{adblockers\xspace}
\providecommand{\adblocking}{adblocking\xspace}
\providecommand{\antiadblocker}{anti-adblocker\xspace}
\providecommand{\antiadblockers}{anti-adblockers\xspace}
\providecommand{\antiadblocking}{anti-adblocking\xspace}
\providecommand{\adblockplus}{AdBlock Plus\xspace}
\providecommand{\ghostery}{Ghostery\xspace}
\providecommand{\privacybadger}{Privacy Badger\xspace}

\providecommand{\vendor}{supplier\xspace}
\providecommand{\vendors}{suppliers\xspace}
\def\top5k{Top-5K\xspace}
\providecommand{\javascript}{JavaScript\xspace}
\providecommand{\javascripts}{JavaScripts\xspace}
\providecommand{\js}{JS\xspace}

\providecommand{\nscripts}{14\xspace}

\providecommand{\vs}{vs.\ }
\providecommand{\ie}{{i.e.}, }
\providecommand{\eg}{{e.g.}, }

\providecommand{\etal}{\emph{et al.}\xspace}

\PassOptionsToPackage{hyphens}{url}
\expandafter\def\expandafter\UrlBreaks\expandafter{\UrlBreaks
  \do\a\do\b\do\c\do\d\do\e\do\f\do\g\do\h\do\i\do\j%
  \do\k\do\l\do\m\do\n\do\o\do\p\do\q\do\r\do\s\do\t%
  \do\u\do\v\do\w\do\x\do\y\do\z\do\A\do\B\do\C\do\D%
  \do\E\do\F\do\G\do\H\do\I\do\J\do\K\do\L\do\M\do\N%
  \do\O\do\P\do\Q\do\R\do\S\do\T\do\U\do\V\do\W\do\X%
  \do\Y\do\Z}

\providecommand{\myparab}[1]{\smallskip\noindent\textbf{#1}}

\newcommand{\squishenum}{   \begin{enumerate}{}    { \setlength{\itemsep}{0pt}
\setlength{\parsep}{0pt}      \setlength{\topsep}{3pt}
\setlength{\partopsep}{0pt}      \setlength{\leftmargin}{1.5em}
\setlength{\labelwidth}{1em}      \setlength{\labelsep}{0.5em} } }
\newcommand{\squishlist}{   \begin{list}{$\bullet$}    {
\setlength{\itemsep}{0pt}      \setlength{\parsep}{3pt}
\setlength{\topsep}{3pt}       \setlength{\partopsep}{0pt}
\setlength{\leftmargin}{1.5em} \setlength{\labelwidth}{1em}
\setlength{\labelsep}{0.5em} } }
\newcommand{\squishlisttwo}{   \begin{list}{$\bullet$}    {
\setlength{\itemsep}{0pt}    \setlength{\parsep}{0pt}
\setlength{\topsep}{0pt}     \setlength{\partopsep}{0pt}
\setlength{\leftmargin}{2em} \setlength{\labelwidth}{1.5em}
\setlength{\labelsep}{0.5em} } }
\newcommand{\squishend}{    \end{list}  }
\newcommand{\squishenumend}{    \end{enumerate} }

\definecolor{red}{RGB}{176,23,31}
\definecolor{green}{RGB}{0,139,69}
\newcommand{\tick}{\color{green}\ding{51}} 
\newcommand{\cross}{\color{red}\ding{55}}
\def\yes{\tick}
\def\no{\cross}

\makeatletter
\def\url@leostyle{%
  \@ifundefined{selectfont}{\def\UrlFont{\small\rmfamily}}%
  {\def\UrlFont{\rmfamily}}%
}
\makeatother	
\definecolor{darkgreen}{RGB}{47,109,79}
\definecolor{darkblue}{RGB}{57,79,99}
\usepackage[bookmarks=false, hidelinks, breaklinks]{hyperref}

\usepackage[hang,flushmargin,stable]{footmisc}

\usepackage{titlesec}
\usepackage{etoolbox}
 \titlespacing\section{0pt}{7pt}{4pt}
 \titlespacing\subsection{0pt}{7pt}{3pt}
 \titlespacing\subsubsection{0pt}{4pt}{3pt}

\begin{document}

\title{\Large \bf Adblocking and Counter-Blocking: A Slice of the Arms Race}

\author{Rishab Nithyanand$^1$, Sheharbano Khattak$^2$, Mobin Javed$^{3}$,\\
Narseo Vallina-Rodriguez$^{4}$, Marjan Falahrastegar$^{5}$, Julia E. Powles$^2$,\\
Emiliano De Cristofaro$^{6}$, Hamed Haddadi$^{5}$, Steven J. Murdoch$^{6}$\\[1.5ex]
{\normalsize $^{1~}$Stony Brook University $\;\;$
$^{2~}$University of Cambridge $\;\;$
$^{3~}$University of California - Berkeley}\\
{\normalsize $^{4~}$International Computer Science Institute $\;\;$
$^{5~}$Queen Mary University of London $\;\;$
$^{6~}$University College London}\\[4ex]}
\date{}

\maketitle

\thispagestyle{empty}

\begin{abstract}

Adblocking tools like Adblock Plus continue to rise in popularity, potentially
threatening the dynamics of advertising revenue streams. In response, a number of publishers
have ramped up efforts to develop and deploy mechanisms for  
detecting and/or counter-blocking \adblockers (which we refer to as {\em
\antiadblockers}), effectively escalating the online advertising arms
race. In this paper, we develop a scalable 
approach for identifying third-party services shared across
multiple websites and use it to provide a first characterization of
\antiadblocking across the Alexa~\top5k websites.  We map
websites that perform \antiadblocking as well as the entities that
provide \antiadblocking scripts. We study the modus operandi of these
scripts and their impact on popular \adblockers. We find that at least
6.7\% of websites in the Alexa~\top5k use \antiadblocking scripts,
acquired from 12 distinct entities -- some of which have a direct
interest in nourishing the online advertising industry.

 \end{abstract}

\section{Introduction} 

\label{intro}

Today's web ecosystem is largely driven by online advertising.
However, recent years have seen a large number of users turn to
\adblocking and tracker-blocking
tools\footnote{While \adblocking differs from
tracker-blocking, to ease presentation, we refer to tools that provide
any of these properties as \adblockers.} for the purposes of improving
their web-browsing experience, maintaining privacy, and more recently
to protect themselves against malware~\cite{abp-most-popular,adblock-malvertiser}.
With a recent study estimating the number of active adblock users to be
198M and revenue losses due to \adblockers at
\$22B~\cite{PageFair-2015}, the threat posed by \adblockers to the
online advertising revenue model has moved from mildly concerning to
existential.  In response, publishers have started to actively detect
users of \adblockers, and subsequently block them or otherwise coerce
them to disable the \adblocker ~-- in the rest of the paper, we refer
to these practices as \emph{\antiadblocking}.  Most recently, this
practice gained wide attention with the endorsement of the Internet
Advertising Bureau (IAB) when, in March 2016, it released a primer on
how to deal with users of \adblockers, as well as a semi open-source
script\footnote{The script was only made available to members of the
IAB.} for detecting the use of
\adblockers~\cite{IAB-adblock-primer:2016}. The tension between key
stakeholders in this ecosystem -- publishers, users, and a plethora of
intermediate beneficiaries  -- %
forms part of what
has been dubbed as the \emph{\adblocking arms
race}~\cite{news-adblocking-wars:2016}.

\myparab{Motivation.} While incidents of
\antiadblocking~\cite{adblockplus-forum:2016,
alexander-hanff-repo:2016,news-adblocking-wars:2016,schneier-adblock:2016},
and the legality of such
practices~\cite{alexander-hanff:2016,adblock-success-law,adblock-fail-law},
have received increasing attention, current understanding thereof is limited to a few
forums~\cite{adblockplus-forum:2016} and user-generated
reports~\cite{alexander-hanff-repo:2016}. As a result, we lack
quantifiable insights into key questions such as: how prevalent
nowadays are such practices on the Web?  Are certain categories of
websites more likely to employ \antiadblockers?  Who are the main
\vendors of \antiadblocking services? What mechanisms do these employ
to detect the presence of \adblockers? Is it possible for \adblockers
to counter-block \antiadblockers? What are common responses
after positive detection of \adblockers and their impact on end-users? In this work, we address
these questions by presenting the first characterization of
\antiadblocking.

\myparab{Roadmap.} We start with characterizing \antiadblocking on the
Web by identifying \antiadblocking scripts across Alexa~\top5k sites.
To this end, we develop a scalable technique to identify popular
third-party services that are shared across multiple websites, and
utilize it to flag \antiadblocking scripts. We then map out the
entities that serve \antiadblocking scripts and the websites that use
these scripts. We find that at least 6.7\% of Alexa~\top5k websites
conduct some form of \antiadblocking by downloading \nscripts scripts
from 12 unique domains most of which belong to ad services, while one
specifically offers \antiadblocking services. Most of the
\antiadblocking websites represent popular categories such as news,
blogs, and entertainment.  We manually visit sample websites from the
\antiadblockers and find that the arms race has already entered the
next round: at least one of three popular browser extensions
(\adblockplus, \ghostery, \privacybadger) can counter-block half of
the \antiadblocking scripts.  We conclude with a discussion of the
\antiadblocking arms race in terms of ethics and legality, also
enumerating existing proposals that aim to achieve a sustainable and
unintrusive online advertising model.

\section{Related Work}

\label{sec:related}

Rafique~\etal~\cite{Rafique:2016} measure \antiadblocking as an incidental aspect of a
broader study of malicious and deceptive advertisements, malware and
scams on {\em free live-streaming services}. They find that \antiadblocking
scripts were used by 16.3\% of the 1,000 domains
they crawled, which is a bit higher than what we find in the Alexa~\top5k (6.7\%),
although not surprising given their heavy use of deceptive ads. 
 
Our paper also complements work quantifying and
characterizing non-transparent third-party web services, as well as
revealing users' differential treatment. For example, Ikram \etal
\cite{ikram2016towards} proposed a machine learning approach to characterize
\javascripts used for online tracking and those used for providing website
functionality. Their work allows privacy-enhancing tools to more selectively
block \javascripts without breaking website functionality.
Acar~\etal~\cite{Acar:2014} and Liu~\etal~\cite{Liu:2013} measure the
prevalence of tracking across large datasets of websites, while
Mayer~\cite{Mayer:2011} studies the effectiveness of some \adblocking
and anti-tracking tools against those sites. Khattak~\etal~\cite{Khattak:2016} assess
discrimination against Tor users at the network and application
layer. Various studies investigate price
discrimination~\cite{Mikians:2013, Hannak:2014}
 and its methods~\cite{Chen:2015} employed by online
marketplaces, and there are other studies on \emph{filter bubbles} --
the effect where high web personalization leads to users being locked
in information silos~\cite{Hannak:2013, Xing:2014}. 

All of these studies illuminate the nature and scale of opaque practices
on the web, informing our understanding of complex and
multidimensional ecosystems. Our work complements previous studies by
presenting a novel technique to identify shared objects across
multiple websites at scale, and utilizing this approach to provide a first look
at how the Web employs \antiadblocking techniques. %

\section{Methodology}

\label{measure}

This section presents our method for identifying third-party services
that are shared between multiple websites. We describe the technique
in the context of identifying shared \antiadblocking \javascripts
(\js). The premise of our approach is that by discovering
\emph{similar} objects (in our case, \javascripts) that are loaded by
multiple websites, we can infer the presence of a common 
third-party \js, its functionality and its source.

\myparab{Crawler overview.} We rely on a Selenium-based web crawler 
to generate the set of \javascripts to analyze. We
load each website in our dataset with four browser modes -- vanilla
Firefox (with no extensions), Firefox with AbBlock Plus, Firefox with Ghostery,
and Firefox with Privacy Badger. For each page load, we capture screenshots,
HTML source code, and responses to all requests generated by the browser.
We extract all the text between \texttt{<script>} and
\texttt{</script>} tags from the HTML and label them as {\em embedded} \js.
Similarly, we detect all \js objects in the collected responses and
label them as {\em downloaded} \js. In total, the~\top5k Alexa websites
generate over 200K individual \js files when loaded with the vanilla Firefox
browser.

\myparab{Identifying \js objects with common sources.}
We formulate our problem of finding groups of similar \js as a maximal
clique finding problem \cite{Bron:1973}. We consider each \js file
loaded by a website to be a node in a graph. If two nodes are within
some margin of similarity of each other  (we define our similarity
metric below), we say there is an edge between them. We extract
classes of \js that have a common source by identifying all maximal
cliques in this graph. By intentionally focusing on finding similar
\js (rather than identical \js) we allow for the grouping of objects
that differ only slightly because they contain website-specific
identifiers, features and properties. 

\myparab{Choice of similarity metric and threshold.}  
In order to add an edge between two nodes in the graph (\ie to indicate that two
\js files in two different websites 
are similar), we need to define a metric for similarity, and a
suitable threshold for this metric. To measure the \emph{similarity} of two \js
files, we use Term Frequency--Inverse Document Frequency (TF-IDF) 
to generate a vector of \emph{keyword weights} for each \js file after
filtering out \js reserved words, such as \texttt{function} and \texttt{var}.
We then use the cosine similarity metric to measure the similarity of the two
keyword weight vectors. Similar approaches using both TF-IDF and cosine
similarity have been used by the information retrieval community for topic
identification and similarity checking of source-code
\cite{Kuhn:2007,Yamamoto:2005}.
We note that this method is particularly well suited to our task 
compared to other string matching approaches because it is:  

\begin{itemize}%
\item {\em White-space insensitive:} Many websites perform script 
minification using different libraries, yielding different 
indentation and white-spacing practices. Our approach is unaffected 
by these complications.
\item {\em Position insensitive:} In scripts that have several 
functionalities (\eg tracking and ad-block detection), the position 
of each specific function is irrelevant to the similarity score.
\item {\em Reasonably resistant to noise:} Small changes (\eg website 
specific identifiers) have little impact on the final similarity score. 
\end{itemize}

\begin{figure}[t]
\includegraphics[trim=0cm 0.1cm 0cm 2.25cm, clip=true,width=.495\textwidth]
{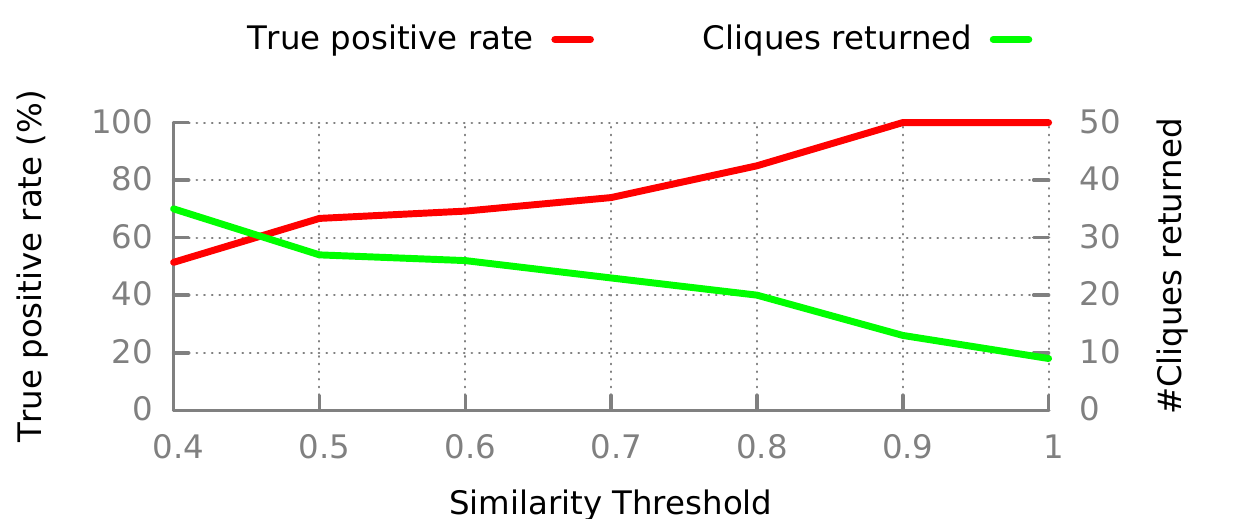}
\vspace{-0.6cm}
\caption{Effect of the similarity threshold parameter on the True Positive Rate
(TPR) and the number of maximal cliques.} %
\vspace{-0.2cm}
\label{fig:threshold-validation}
\end{figure}

In order to determine a similarity score \emph{threshold}, we perform a series 
of experiments on a small dataset of 4.4K \js files extracted 
from the Alexa Top-100 websites. In each experiment, we set a similarity
threshold between $0.40$ and $1.00$ and compute the cliques in each of the
corresponding graphs. We then manually inspect the cliques extracted at each
threshold to identify the fraction of cliques containing \js with
identical functionality and sources. Using this approach, we find that at a 
similarity threshold of $0.80$, 17/20 cliques returned by our program contain
scripts  with identical functionality and sources, \ie achieving True Positive
Rate (TPR) of 0.85. In Figure~\ref{fig:threshold-validation}, we illustrate the 
change in TPR along with the number of cliques returned as the threshold 
increases.  Although thresholds above 0.90 yield TPR$=$1.0, the number of 
cliques returned drops significantly, which will result in lower True
Negative Rates (TNR). Therefore, following a conservative stance, we
use a threshold of 0.80 for the remainder of our experiments.

\myparab{Improving scalability.} Our approach involves computing the cosine
similarity between each pair of keyword weight vectors, thus requiring $O(n^2)$
vector multiplications for $n$ \js files. Given the large number of
\js files used by websites (\eg the Alexa~\top5k sites contained over
200K \javascripts), this may not scale with large datasets. Therefore,
we use a set of heuristically developed filters to eliminate
comparisons between scripts that are unlikely to ever be part of the
same clique:

\begin{itemize}%

\item \emph{Word-count filter:}  We avoid comparing scripts with significant
word-count difference. Specifically, if a pair of scripts has a word-count ratio
higher than 1.50, we assume that they are unlikely to be a part of the same
clique and set their similarity to 0.

\item \emph{Embedded \vs downloaded script filter:} \javascript is either
embedded in the source HTML for page-specific functionality, or downloaded
separately from external sources to provide site-wide functionality. We
do not consider them as the same type of identity thus we set their
similarity to 0.

\item \emph{Source filter:} If two \javascripts are fetched from
the exact same URL, we mark them as identical. 

\item \emph{\js domain filter:} \javascript can communicate with
external sources indicated by embedded URLs. We assume that for any
pair of scripts, if one communicates with external sources and the
other does not, their functionality is different and set their
similarity score to 0.

\end{itemize}

\myparab{Source and functionality identification.}  Once maximal cliques of
similar scripts are identified, the content and meta-data of each script in a
clique is used to generate and log: \first the FQDN (Fully Qualified Domain
Name) of the script's source, \second FQDNs of external resources utilized by
the script, and \third keywords associated with the script. In
Section~\ref{result}, we use these three features, in addition to content of the
script, to classify cliques by functionality.

\myparab{Method limitations.} We acknowledge that our method has a
few limitations. First, our similarity metric will fail to
identify obfuscated \js code. Second, given that we do not
compare downloaded with embedded \js code, we may fail to identify
small cliques in which a reduced number of sites integrate an
\antiadblocking \js in a different way than is normal. Finally,
our method may fail to identify similarities between composed \js --
\ie scripts that consist of multiple individual files downloaded as a
single object. As a result, our method only provides a lower-bound
approximation of the usage of \antiadblocking across websites. We plan
on addressing these limitations in future work.

\section{Dataset and Results}
\label{result}

\begin{table}
\footnotesize
	\begin{center}
	\begin{tabular}{lrr}
	\toprule
		&	{\bf Cliques} & {\bf Websites}  \\
		\midrule
		{\bf Downloaded} & 1,373 & 3,619 \\
		{\bf Embedded} & 509 & 2,070 \\
		{\bf Trackers} & 456 & 2,741 \\
		{\bf Anti-Adblockers} & 22 & 335 \\
    \bottomrule
    \end{tabular}
	  \vspace{-0.1cm}
		\caption{The number of total cliques (out of 1,882 found) and those related to tracking and \antiadblocking, along with the number of websites that incorporate these scripts (totalling 4,017 websites, computed over 200K downloaded and embedded scripts).~\nv{A row with the total \# of js embedded and downloaded wouldn't harm}}
	\label{table:tag}
	  \vspace{-0.6cm}
	\end{center}
\end{table}

We apply our clique detection methodology to the \js objects fetched
by our crawler using the vanilla Firefox browser. We restrict our analysis to
cliques of size greater
than 5 --~\ie \javascripts shared by more than 5 sites in our dataset
-- as we are interested in identifying scripts that are shared across
many websites. We acknowledge that this approach might fail to
flag \antiadblocking scripts utilized by individual or a small number
of websites, and those used by a few websites in the Alexa~\top5k but
popular among websites ranked above 5K. As shown in
Table~\ref{table:tag}, we find 1,373 cliques that are shared among
3,619 websites in the downloaded files, with an average of 232
websites per clique ($\sigma$=365.6) and the largest clique having
1,320 websites (which we find, via manual inspection, is a \js related
to jQuery). Among the embedded scripts, 509 cliques are
shared by 2,070 websites ($\mu$=41.2 $\sigma$=48.9 max=261). 

We manually analyze all the 1,882 cliques (corresponding to 4,017 unique
websites) identified for both downloaded and embedded scripts, and tag
them as \emph{trackers} (if they upload information such as IP
addresses and cookies to tracking companies), \emph{\antiadblockers}
(if they check for the presence of \adblockers), or \emph{others}. Manual
analysis is performed by identifying external libraries and function specific
keywords used in the scripts. We
note that manual analysis of \js is a tedious process that does not
scale to a larger number of scripts, therefore we leave as part of future work
to investigate ways  to automate \js tagging. 
 
We uncover 22 cliques used for \antiadblocking employed by 335
websites -- about 6.7\% of Alexa~\top5k websites. We observe that
Alexa Top-1K have 60 \antiadblocking websites, and the number
increases by about 70 websites for every additional 1K considered,
reaching 335 \antiadblocking websites in~\top5k.\nv{That would be
great in a table/figure if we have real state} \edc{perhaps
CDF?}\sbk{We don't have space for this. If agree, please remove
comments.} While studying \antiadblockers, we also identify 456
tracking cliques employed by about 54\% of Alexa~\top5k, validating
previous studies on the pervasiveness of tracking over the
Web~\cite{Marjan:2016}.

\myparab{Anti-adblocking by website categories.} In
Table~\ref{table:categories}, we report the categories of the 335
\antiadblocking websites, using McAfee's URL categorization
service~\cite{mcafee-categ}. We find that \antiadblocking is common
among a diverse mix of publishers, and prevalent among publishers of
``General News'' (19.5\%), ``Blogs/Wiki'' (9.3\%), and
``Entertainment'' (8.5\%) categories, which represent more than one
third of all websites. Note that these categories are also among the
most popular ones across all~\top5k Alexa domains, although to a
lesser extent -- respectively, 9.4\%, 6.29\%, and 5.4\%.  Whereas,
other popular categories among~\top5k domains (e.g., ``Internet
services'', ``Online Shopping'', ``Business'', which account for 20\%
of the~\top5k) are much less prevalent in \antiadblocking websites.

\begin{table}
\small
  \begin{center}
    \resizebox{\columnwidth}{!}{  
  \begin{tabular}{rl|rl}
  \toprule
  {\bf \%} & {\bf Category} &   {\bf \%} & {\bf Category}\\
  \midrule
  19.5\% & General News &  2.5\% & Pornography \\
  9.3\% & Blogs/Wiki & 2.5\% & Forum/Bulletin Boards \\
  8.5\% & Entertainment & 2.2\% & Technical/Business Forums \\
  4.3\% & Internet Services  & 2.2\% & Potential Illegal Software \\
  3.7\% & Sports &  2.0\% & Online Shopping \\
  3.7\% & Games &  1.7\% & Portal Sites \\
  3.2\% & Travel &  1.7\% & Humor/Comics \\
  3.2\% & Education/Reference  & 1.2\% & Social Networking \\
  2.7\% & Business &  1.2\% & Provocative Attire \\
  2.5\% & Software/Hardware &  1.2\% & Marketing/Merchandising \\
  \bottomrule
\end{tabular}
}
  \vspace{-0.2cm}
\caption{\small Distribution of \antiadblocking websites
by category according to McAfee's URL categorization.}
  \label{table:categories}
  \vspace{-0.6cm}
  \end{center}
 \end{table}

\myparab{Website response to detection of \adblockers.} In order to
assess how \antiadblocking websites behave once they identify
\adblockers, we look at all the screenshots taken by our crawler,
respectively, when using the vanilla Firefox browser with no extensions and the
Firefox browser with AdBlock Plus enabled (which we assume is more likely to be
detected due to its popularity~\cite{abp-most-popular}) \edc{move/clarify this
in methodology section?}\sbk{We say in crawler overview in methodology
that we take screenshots for all popular adblockers, but here we are
manually looking at screenshots of adblockplus only; we look at all
websites for adblock plus and then look at screenshots w/ other
adblockers when there's some response to adblockplus. Please remove
comments if satisfied (or edit as suitable)}.

We note cases where there is an explicit (\ie warning to disable
\adblocker) or a discrete (\ie blank page via AdBlock Plus, but normal
appearance without) response to \adblocking.  For these websites, we
also view screenshots when accessed by the Firefox browser with each of the
following extensions: Ghostery, Privacy Badger, and NoScript.

We find only 6 explicit and no discrete responses to \adblocking.  Of
the explicit responses, 3 are displayed by porn websites hosted by the
same company -- \texttt{MindGeek} -- and employ the same
\antiadblocking script downloaded from \texttt{DoublePimp}.  The
warning is displayed for both AdBlock Plus and Ghostery.  The
remaining 3 also employ the same script, but display different
messages (only for AdBlock Plus) with the same general theme,~\ie
nudging the user to disable the \adblocker and/or support the website
via subscription or donation. 

Some websites display \adblocker warning to users after they engage in
some form of activity, such as clicking on links or scrolling. To
capture such responses, we repeat the above exercise for screenshots
taken after mimicking user activity -- specifically, clicking on a
random link on the page, scrolling down to the bottom of the newly
loaded page, waiting three seconds, then scrolling back up to the top
of the page, waiting 5 seconds. While the modified methodology
validates our previous observations, we do not discover any new
responses. 

In the attempt of automating the analysis of websites' response to
\antiadblocking, we have also tried
to use image comparison tools, such as perceptual
hashing. %
However, this generates
a  high number of false positives due to dynamic content on many
sites %
as well as false negatives since \antiadblocking warnings and messages generate a
relatively small visual difference.

\myparab{How \antiadblockers work.} Next, we manually inspect the 22
\antiadblocking scripts (14 downloaded and 8 embedded) aiming to
understand how \antiadblocking scripts detect \adblockers. We note
that of these only the 14 downloaded scripts are actually useful as
the 8 embedded scripts simply redirect to the downloaded scripts. We
find that \antiadblockers operate on a simple premise: if a bait object (\ie an
object that is expected to be blocked by ad-blockers -- \eg a \js or
\texttt{DIV} element named \texttt{ads}) on the publisher's website is missing
when the page loads, the script concludes that the user has an
\adblocker installed.

Specifically, the \antiadblocker detects \adblockers by one of the following
approaches: (1) The \antiadblocker injects a bait advertisement container
element (\eg \texttt{DIV}), and then compares the values of properties
representing dimensions (\texttt{height}  and \texttt{width}) and/or visual
status (\texttt{display}) of the container element with the expected values when
properly loaded. (2) The \antiadblocker loads a bait script that modifies the
value of a variable, and then checks the value of this variable in the main
\antiadblocking script to verify that the bait script was properly loaded. If
the bait object is determined to be absent, the \antiadblocking script concludes
that an \adblocker is present.
To track whether the user has turned off the \adblocker after being
prompted to do so, the \antiadblocker periodically runs the ad-block
check and stores the last recorded status in the user's browser using
a cookie or local storage.  

\begin{table}
\small
\setlength{\tabcolsep}{2pt}
        \begin{center}
        \begin{tabular}{l l r | c c c}
	\toprule
     {\bf Domain } & {\bf Description} & {\bf \#Sites} & \textbf{ABP} & \textbf{Gh} & \textbf{PB} \\
\midrule
pagefair.com & Anti-adblocking & 20 & \yes & \no & \yes \\
googleadservices.com   & Ads & 61      &       \no     &       \no     &       \no     \\
googlesyndication.com & Ads & 13 & \no &\no & \no \\
taboola.com  &  Ads &     36 &  \no     &       \yes    &       \yes    \\
outbrain.com  & Ads         & 10     &       \no     &       \yes    &       \yes    \\
ensighten.com       & Ads & 6 &       \no     &       \yes    &       \no     \\
hotjar.com     & Analytics     &     9 &  \no     &       \no     &       \no     \\
doublepimp.com  & Pornography    &   8 &    \no     &       \yes    &       \no     \\
tacdn.com    & Travel &    8 &   \no     &       \no     &       \no     \\
cloudflare.com    & CDN  & 50 &      \no     &       \no     &       \yes    \\
cloudfront.net     &  CDN  &  6 &  \no     &       \no     &       \no     \\
ytimg.com & Content/Ads & 108 & \no & \no & \no \\
\bottomrule
    \end{tabular}
    \vspace{-0.25cm}
    \caption{\small Domains from which \antiadblocking scripts are downloaded and \#websites employing them. The table's right side reports whether AdBlock Plus, Ghostery, and Privacy Badger counter-block \antiadblocking scripts from these domains.}
        \label{table:adb_tools}
            \vspace{-0.6cm}
        \end{center}
\end{table}

\myparab{Anti-adblocker \vendors.} We analyze the source code of the 14
\antiadblocking scripts and the domains from which these are
downloaded aiming to infer the \vendors of these scripts. 
The remaining 8 embedded scripts redirect to \antiadblocking scripts served by
\texttt{Cloudflare} and \texttt{Taboola}.\nv{Can we say why
cloudflare provides this technology if this is not part of their
business? }\sbk{In the next para, we say that we can't say who is the
real \vendor for scripts serverd via CDNs. Please remove comments if
satisfied.} Our analysis is summarized in Table~\ref{table:adb_tools}.
We also include a description of these domains  -- based on the
information available on their official websites, Google search, and
McAfee URL categorization service~\cite{mcafee-categ} -- as well as
the number of websites in our dataset that employ the \antiadblocker. 

At the top we find \texttt{Pagefair}, a company
specialized in \antiadblocking services, followed by a number of
domains related to \texttt{Google}, \texttt{Taboola},
\texttt{Outbrain} and \texttt{Ensighten}. Overall the \antiadblockers
downloaded from these 5 domains are employed by 48\% of all the 315
websites employing \antiadblockers. We note that these domains are
direct beneficiaries of \antiadblocking as these inherently thrive on
the prevalence of online advertisements. Though not directly related
to online advertisement, the ability to detect \adblockers is a useful
capability for the analytics company
\texttt{HotJar}. %

We also find two
cases where the \antiadblocking script is shared by entities in the
same domain or business: \texttt{TripAdvisor} (\emph{tacdn.com})
distributes the script to its 8 websites with different country code
top-level domains. Adult websites, all of which are hosted by
\texttt{MindGeek}, turn to \texttt{DoublePimp} for \antiadblocking.
Two \antiadblocking scripts are pulled from popular Content Delivery
Networks (CDNs), but we could not determine their original \vendor.
Finally, \texttt{ytimg} (a content server associated with YouTube)
serves a script that has the ability to detect if ads were properly
loaded, however, it is not clear how it uses this information.

\myparab{Adblocker response to being blocked.} There is anecdotal
evidence that the \adblocking arms race has entered the next level:
some \adblockers can detect \antiadblockers and counter-block
them~\cite{counter-blocking-news}.  To test for this behaviour, we
visit a sample website for each \antiadblocking script via
\adblockplus, \ghostery and \privacybadger over Chrome web browser.
We repeat the experiment three times and monitor all HTTP requests
generated when loading the website using Chrome's \emph{Developer
Tools}. We infer that \adblocker can counter-block if the request to
fetch \antiadblocking script fails to be initiated. As reported in
Table~\ref{table:adb_tools}, half of the 12 \antiadblocking \vendors
are blocked by at least one \adblocker.  \ghostery and \privacybadger
detect 4 \antiadblockers each, while \adblockplus detects only 1.
Anti-adblocking scripts served by \texttt{Taboola} and
\texttt{Outbrain} are blocked by both \ghostery and \privacybadger,
\texttt{PageFair} scripts by both \adblockplus and \privacybadger,
while \texttt{Doublepimp}, \texttt{Ensighten} and \texttt{Cloudflare}
scripts by at most one of the three \adblockers.  We note that the
\antiadblocking \vendors that are never detected are related to
content distribution, Google ad services, analytics, or site-wide
scripts. 

\section{Discussion}

\label{discussion}

The \adblocking arms race involves a plethora of players: between
publishers and consumers, a jostling array of intermediaries compete
to deliver ads, mostly supported by business models that involve
taking a cut of the resultant advertising revenue. At the heart of
this rich ecosystem lie important questions regarding the legality and
ethics of \adblocking and \antiadblocking.

The legality of \adblocking is potentially contestable under laws about anti-competitive business conduct and copyright infringement. To date, only Germany has tested these arguments in court, with \adblockers winning most~\cite{adblock-success-law}, but not all of the cases~\cite{adblock-fail-law}. On the other hand, \antiadblocking in the EU might in turn breach Article 5(3) of the Privacy and Electronic Communications Directive 2002/58/EC, as it involves interrogating an end-user's terminal equipment without consent~\cite{alexander-hanff:2016}.
~\nv{It would be nice if we have something about the tracking activities of Anti-adblockers in S4}

Many consider \adblocking to be an ethical choice for consumers and
publishers to consider from both an individual and societal
perspective. In reality, however, both sides have resorted to radical
measures to achieve their goals. The Web has empowered publishers and
advertisers to track, profile and target users in a way that is
unprecedented in the physical realm~\cite{Marjan:2016}. In addition,
publishers are inadvertantly and increasingly serving up malicious 
ads~\cite{adblock-malvertiser}. This has resulted in the
rise of \adblocking, which in turn has led publishers to employ
\antiadblocking. The core issue is to get the balance right between
ads and information: publishers turn to \antiadblocking to force
consumers to reconsider the default blocking of ads for earnest
ad-supported publishers but defaults are difficult to shift at
scale. Nevertheless, those publishers will fail if they do not redress
in a fundamental way the reasons that brought consumers to \adblockers
in the first place. There exist proposals to provide a compromise,
such as privacy-friendly advertising~\cite{Guha:2011} as well as
mechanisms to give users more control over ads and trackers they are
exposed to~\cite{Yu:2016, Achara:2016}. Our work extends these efforts
by providing quantified insights into \antiadblocking, to inform
policy that can improve upon the current blocking/counter-blocking
deadlock.

\section{Conclusion}

\label{conclusion}

This paper presented a measurement-based analysis aimed to provide a
first look at the arms race between adblocking and \antiadblocking.  We
found that at least 6.7\% of Alexa~\top5k websites, mostly in popular
categories like news, blogs, and entertainment, engage in some form of
\antiadblocking. The arms race has already entered the next level, as
at least one of three popular browser extensions -- \adblockplus,
\ghostery, \privacybadger\ -- can evade half of the \antiadblocking
scripts in our dataset. In future work, we plan to extend our
measurements beyond the Alexa~\top5k websites, and experiment with
crowdsourced and/or automated mechanisms to tag \javascript by
functionality and to assess publisher response to detection of
\adblockers.

\myparab{Acknowledgements.}
The authors would like to thank the anonymous reviewers for constructive
feedback on preparation of the final version of this paper. 
Rishab Nithyanand was supported by a Open Technology Fund Emerging Technology
Senior Fellowship. Sheharbano Khattak and Steven J. Murdoch were supported by
The Royal Society [grant number UF110392]; Engineering and Physical Sciences
Research  Council [grant number EP/L003406/1]. Emiliano De Cristofaro was
supported by a Xerox University Affairs Committee award and EU grant
H2020-MSCA-RISE ``ENCASE''. 

\myparab{Source code and data release.}
The source code of our \js clique extraction approach can be found at
\url{https://bitbucket.org/rishabn/ad-study-code}. Data created during this
research is available from the University of Cambridge data archive at
\url{http://dx.doi.org/10.17863/CAM.703}. 

\balance
\small
\bibliographystyle{abbrv}
\bibliography{paper}

\end{document}